# Channel-Adaptive Sensing Strategy for Cognitive Radio Ad Hoc Networks

Yuan Lu, Alexandra Duel-Hallen
Department of Electrical and Computer Engineering
North Carolina State University
Raleigh, NC, 27606
{ylu8, sasha}@ncsu.edu

*Abstract*—*In Cognitive Radio (CR) ad hoc networks, secondary users (SU) attempt to utilize valuable spectral resources without causing significant interference to licensed primary users (PU). While there is a large body of research on spectrum opportunity detection, exploitation, and adaptive transmission in CR, most existing approaches focus only on avoiding PU activity when making sensing decisions. Since the myopic sensing strategy results in congestion and poor throughput, several collision-avoidance sensing approaches were investigated in the literature. However, they provide limited improvement. A channel-aware myopic sensing strategy that adapts the reward to the fading channel state information (CSI) of the SU link is proposed. This CSI varies over the CR spectrum and from one SU pair to another due to multipath and shadow fading, thus randomizing sensing decisions and increasing the network throughput. The proposed joint CSI adaptation at the medium access control (MAC) and physical layers provides large throughput gain over randomized sensing strategies and/or conventional adaptive transmission methods. The performance of the proposed CSI-aided sensing strategy is validated for practical network scenarios and demonstrated to be robust to CSI mismatch, sensing errors, and spatial channel correlation.*

I. INTRODUCTION

In Cognitive Radio (CR) networks, secondary users (SU) attempt to communicate over a set of channels without causing significant interference to the primary users (PU) [1-3]. Many researchers have investigated spectrum sensing and access under the hardware constraints on sensing capability [1-4]. These methods can potentially offer a solution to rapidly growing spectrum crunch, but there are significant hurdles to overcome before commercial deployment of CR techniques becomes feasible. Exploitation of fading channel state information (CSI) is a promising approach to throughput improvement in CR systems. The focus of this paper is CSI-aware sensing and access in mobile CR ad hoc networks. *The main premise is that it is feasible and beneficial for SUs to anticipate upcoming channel fades and shadowing (in addition to PU availability) and to adapt to these parameters when sensing and selecting channels for packet transmission while protecting the PUs.*

Adaptive transmission in spectrum sharing environments has also received considerable attention, e.g., [5,6]. These investigations show that network throughput can be boosted by exploiting the instantaneous CSI between network nodes. However, in the proposed systems the adaptation to fading CSI only takes place during transmission, but not in the MAC layer.

Sensing strategies in the literature, e.g., [1,4,7-11] fail to exploit the underlying wireless channel. The myopic sensing strategy [1] makes sensing decisions according to the expected reward given by the channel bandwidth. This strategy results in SU congestion and poor throughput in traffic-clogged networks since most neighboring SUs attempt to sense and transmit over the same channels with low PU activity. Randomized and cooperative strategies address SU congestion by randomizing sensing decisions or employing user negotiation [7-11]. However, these approaches provide limited throughput improvement.

We propose to exploit spatial variation of SU links to resolve competition among SUs and to boost the throughput. The conventional reward choice given by channel bandwidth is identical for all SUs. On the other hand, in the proposed strategy, *the reward is adapted to the maximum achievable rate that depends on the CSI* (i.e., the instantaneous channel gain) of individual SU links. This CSI varies over the locations of SUs and over the CR spectrum due to channel fading, thus providing *multiuser diversity*.

Several practical issues need to be addressed to validate feasibility and performance of the proposed CSI-aided sensing strategy in practical CR networks. First, it requires the knowledge of the channel gain prior to sensing. The SUs can learn this CSI by exchanging low-rate Ultrawideband (UWB) pilot signals under the noise level of the active PUs [12]. Alternatively, a small portion of bandwidth can be dedicated to transmission of low-rate pilots over idle PU channels. Based on these pilots, fading prediction techniques can be employed to compensate for CSI delay [13]. However, CSI mismatch is expected to be more significant than in conventional adaptive transmission approaches. Therefore, we address robustness of the proposed channel-adaptive sensing strategy to CSI estimation errors. Second, we validate performance of CSI-aided sensing for multipath and shadow fading channel models. The channel gain in the latter model is slowly varying and is easier to estimate prior to sensing. However, it is subject to spatial correlation that can potentially reduce the benefit of multiuser diversity. Sensitivity of the proposed sensing approach to spatial correlation is addressed using long-term fading models in the literature [14,15]. Finally, we analyze performance of the proposed sensing approach for practical CR systems, including combined CSI-aided sensing and adaptive modulation and performance under sensing errors.

The rest of this paper is organized as follows. In section II,

This research was supported by the NSF grant CNS-1018447.

we present the system model and describe the proposed CSI-aided sensing strategy. Design and performance issues for realistic CR systems are discussed in section III. Numerical results are provided in section IV. Conclusions and future work are discussed in section V.

## II. CHANNEL-ADAPTIVE MYOPIC SENSING STRATEGY

Consider an overlay CR network with $M$ SU transmitter-receiver pairs and $N$ non-overlapping channels. The PU traffic is modeled as a stationary Markov process with independently evolving channels. The transition probabilities are assumed known to all SUs. For channel $n$ at the $m^{th}$ SU location, $p_{ij}^{mn}$ denotes the probability of transition from state $i$ to state $j$, where $i,j \in \{0\ (busy),\ 1\ (idle)\}$ as shown in Fig. 1. Both the primary network and the CR network share the same slotted structure, and all users (including PUs and SUs) are perfectly synchronized [1].

Assume SUs can sense only one channel at a given time slot due to the hardware constraints. The belief vector $\boldsymbol{\theta}^m(t) = [\theta^{m1}(t), ..., \theta^{mn}(t), ..., \theta^{mN}(t)]$, where $\theta^{mn}(t)$ is the conditional probability that channel $n$ is available at time $t$ at the $m^{th}$ SU location based on past sensing outcomes, is employed by the SUs to infer the current state of the PU traffic [10]. Assume a *myopic*, or greedy, sensing policy [1]. At the first time slot $t=1$, the initial belief vector is given by the stationary probabilities of the Markov process. Then at each time slot $t>1$, SU $m$ chooses to sense the channel $n_*^m(t)$ by maximizing *the expected reward* $E[R^{mn}(t)]$:

$$n_*^m(t) = \arg\max_n \theta^{mn}(t) R^{mn}(t) \quad (1)$$

Assuming no sensing errors, the sensing result $a^m(t) = 1$ if the channel is idle and $a^m(t) = 0$ otherwise.

The belief vector is updated according to the sensing result as

$$\theta^{mn}(t+1) = \begin{cases} p_{11}^{mn}, & \text{if } n_*^m(t)=n, a^m(t)=1 \\ p_{01}^{mn}, & \text{if } n_*^m(t)=n, a^m(t)=0 \\ \theta^{mn}(t)p_{11}^{mn} + (1-\theta^{mn}(t))p_{01}^{mn}, & \text{if } n_*^m(t) \neq n \end{cases} \quad (2)$$

and the process is repeated over the time horizon $t \in [1,T]$.

The myopic policy has good performance [1] and is optimal when all channels have the same transition probabilities [16] in the single SU pair scenario. However, the myopic policy ignores competing SUs, and its performance degrades when multiple SUs are active. The PU transmission range is typically larger than the SU range [2] as shown in Fig. 2. Therefore, neighboring SUs are very likely to be affected by the same set of PUs, and the belief vectors $\boldsymbol{\theta}^m(t)$ of these SUs arrive at similar values as $t$ increases. As a result, neighboring SUs tend to choose the same channel to sense, leading to reduced individual throughput. Moreover, when SUs compete for the same channels, they leave other channels unexploited, thus degrading the network throughput.

One reason why the myopic policy is degraded in traffic-congested scenarios is the conventional reward choice given by

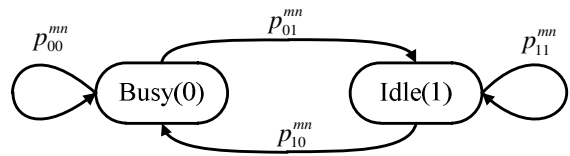

Fig. 1: State transition of PU traffic for channel $n$ at the $m^{th}$ SU.

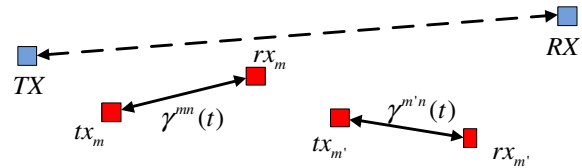

Fig. 2: A PU TX/RX pair and SU tx/rx pairs $m$ and $m'$.

the channel bandwidth

$$R_{conv}^{mn}(t) = B_n, \quad (3)$$

which is often normalized to one. With this reward choice, the most likely channel to be idle in the current time slot is sensed, resulting in poor performance in congested CR networks. To reduce congestion, several strategies in the literature [7-11] randomize sensing decisions or use negotiation to avoid collisions while retaining the reward choice (3). However, the gain of these strategies is limited.

*We propose to adapt the reward to the power of the link between the transmitting and receiving secondary network nodes*. The received powers of the SU links vary in space and frequency due to channel fading. Thus, adaptation of the reward to the channel gain *randomizes sensing decisions and boosts the network throughput*.

Consider the $m^{th}$ SU transmitter-receiver pair. At time slot $t$, the received signal-to-noise ratio (SNR) of the link between these two SUs on the $n^{th}$ channel is given by $\gamma^{mn}(t) = g^{mn}(t)P/N_0 B_n$ as illustrated in Fig. 2, where $g^{mn}(t)$ is the received channel gain (the CSI), $P$ is the transmission power, $N_0$ is the power spectral density of the complex Additive White Gaussian Noise and $B_n$ is the bandwidth of the $n^{th}$ channel. Initially we assume the value of $g^{mn}(t)$ (or, equivalently, the value of $\gamma^{mn}(t)$) is perfectly known at the $m^{th}$ SU transmitter and is fixed for the duration of the slot. In the proposed CSI-aided policy, the reward is given by the channel capacity

$$R_{cap}^{mn}(t) = C^{mn}(t) = B_n \log_2(1+\gamma^{mn}(t)) \quad (4)$$

This reward is a function of the instantaneous CSI and differs among SU pairs, randomizing the sensing decisions and improving the individual throughput.

## III. CSI-AIDED MAC FOR REALISTIC CR NETWORKS

In this section we discuss practical implementation issues of channel-adaptive sensing. First, the reward (4) defines the maximum achievable rate, which is not attainable in practice. Thus, we investigate cross-layer design that combines CSI-aided sensing with *adaptive modulation*. For simplicity we confine our attention to continuous rate adaptation [17] although this approach can be easily extended to discrete-rate adaptation. Assuming adaptive QAM modulation with fixed transmission power, we employ adaptively adjusted data rate as the reward,

$$R_{AM}^{mn}(t) = B_n k^{mn}(t), \quad (5)$$

where $k^{mn}(t)$ is equal to the maximum spectral efficiency that the system can support under a certain BER constraint $BER_T$ [17],

$$k^{mn}(t) = \log_2\left(1 - 1.5\gamma^{mn}(t)/\ln(5BER_T)\right) \quad (6)$$

Second, the proposed sensing strategy relies on the knowledge of the fading channel SNR $\gamma^{mn}(t)$ that might be difficult to estimate accurately since channels cannot be fully explored prior to sensing. Thus, it is important to ensure that the proposed policy is robust to *CSI mismatch*. In the equations below, we suppress the indices $m, n$ and $t$ for brevity. We assume that the sensing policy employs the Minimum Mean Square Error (MMSE) estimate of the actual SNR $\gamma$ conditioned on its mismatched observation $\hat{\gamma}$. Given $\hat{\gamma}$, the reward for each SU link is computed as

$$R(\hat{\gamma}) = \int_0^{+\infty} R(\gamma)f(\gamma|\hat{\gamma})d\gamma \quad (7)$$

where $R(\gamma)$ is given by (4) or (5), and $f(\gamma|\hat{\gamma})$ is the probability density function of $\gamma$ conditioned on $\hat{\gamma}$. For Rayleigh fading channels [18]

$$f(\gamma|\hat{\gamma}) = \frac{1}{\bar{\gamma}\sigma^2} I_0\left(\frac{2\sqrt{\gamma\hat{\gamma}}}{\bar{\gamma}\sigma^2}\right)\exp\left(\frac{-1}{\bar{\gamma}\sigma^2}(\gamma+\hat{\gamma})\right), \quad (8)$$

where $I_0(\cdot)$ is the $0^{th}$ order modified Bessel function, $0 \leq \sigma^2 \leq 1$ is the normalized mean-square-error (NMSE) of the SNR estimation, and $\bar{\gamma}$ is the average SNR.

Finally, to analyze performance of CSI-aided sensing in the presence of *sensing errors*, we employ the belief update process that incorporates the reliability of the spectrum sensor, which is characterized by the miss detection (when the sensor fails to detect the PU signal) rate $p_m$ and the false alarm (when the sensor misidentifies the PU presence) rate $p_f$. Assuming energy detection [19],

$$p_m = 1 - \int_0^\infty Q_\nu(\sqrt{2\nu\lambda}, \sqrt{\tau})f(\lambda)d\lambda \quad (9)$$

$$p_f = \frac{\Gamma(\nu, \frac{\tau}{2})}{\Gamma(\nu)}, \quad (10)$$

where $\nu$ is the number of collected samples, $\tau$ is the detection threshold, $f(\lambda)$ is the distribution of the SNR of the PU signal observed at the sensor, $\Gamma(.)$ and $\Gamma(.,.)$ are complete and incomplete gamma functions, respectively, and $Q_\nu(.,.)$ is the generalized Marcum Q-function. Suppose the node-level constraint [3] on the collision probability between the SU and the PU networks is known. Given $p_m$, the detection threshold $\tau$ is determined by inverting (9), and the false alarm rate $p_f$ is then computed from (10).

The belief vector $\boldsymbol{\theta}_r^m(t)$ is updated according to the sensing result and the sensor reliability information as [20]:

$$\theta_r^{mn}(t) = \begin{cases} \dfrac{(1-p_f)\theta^{mn}(t)}{(1-p_f)\theta^{mn}(t) + p_m(1-\theta^{mn}(t))}, & \text{if } n_m^*(t) = n, a^m(t) = 1 \\[2ex] \dfrac{p_f\theta^{mn}(t)}{p_f\theta^{mn}(t) + (1-p_m)(1-\theta^{mn}(t))}, & \text{if } n_m^*(t) = n, a^m(t) = 0 \\[2ex] \theta^{mn}(t), & \text{if } n_m^*(t) \neq n \end{cases} \quad (11)$$

$$\theta^{mn}(t+1) = p_{mn}^{11}\theta_r^{mn}(t) + p_{mn}^{01}(1-\theta_r^{mn}(t)) \quad (12)$$

## IV. NUMERICAL RESULTS

We assume the CR network traffic is backlogged initially and remains backlogged over the entire time horizon. Consider a CR network with $M = 20$ SU pairs and $N = 40$ channels with the same bandwidth $B = 1$. The PU traffic evolves independently over each channel, and the transition matrix of the channel state is the same for all channels and all SUs,

$$P = \begin{bmatrix} p_{00} & p_{01} \\ p_{10} & p_{11} \end{bmatrix} = \begin{bmatrix} 0.8 & 0.2 \\ 0.2 & 0.8 \end{bmatrix} \quad (13)$$

These assumptions are valid throughout this section unless noted otherwise.

### A. Perfromance of Ideal CSI-Aided Policy

First, suppose all SU links are subject to *independent Rayleigh fading* with the average SNR=10dB. The instantaneous SNR $\gamma$ is fixed over the duration of one time slot and assumed to be known at the transmitter for each channel. The slot duration is 1 ms, and the maximum Doppler shift is $f_{dm} = 40$ Hz.

In Fig. 3, the throughput is compared for five sensing strategies: the random channel selection approach, the conventional myopic policy [1,21], the randomized myopic policy [8], the modified myopic policy with collision avoidance (myopic/CA) from [11], and the proposed approach, the CSI-aided myopic policy with the reward given by (4). Note that for all policies in Fig. 3, the accumulated reward is given by the channel capacity, i.e., they *all utilize adaptive transmission once the channel is sensed*. However, only the CSI-aided strategy adapts to the channel gain when making sensing decisions. The throughput values represent the *average normalized maximum achievable throughput* defined as the network throughput obtained when channel capacity is the accumulated reward divided by the number of SU pairs $M$ in the CR network averaged over the last $t$ time slots.

We assume that an SU will transmit over the channel if it is sensed to be idle or go to sleep during the current time slot if the channel is busy. If multiple SU pairs choose to sense the same channel, and the channel is idle, only one of them can transmit successfully. This can be accomplished by using a carefully designed MAC scheme. In this paper, we employ a slightly modified version of the MAC algorithm in [22].

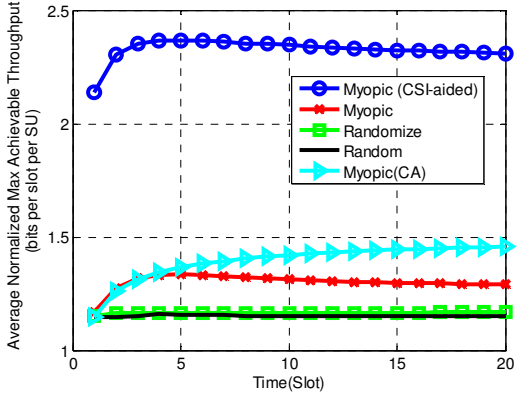

Fig. 3: Throughput vs. time; capacity reward; 20 SU pairs; 40 channels; i.i.d Rayleigh fading; average SNR=10dB.

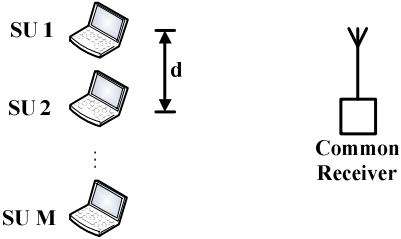

Fig. 4: Network topology for the lognormal shadowing model.

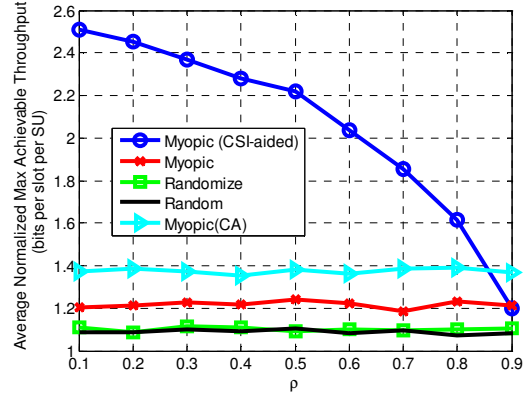

Fig. 5: Throughput vs. spatial correlation $\rho$; capacity reward; 20 SU pairs; 40 channels; log-normal fading; average SNR=10dB; $\sigma_{\gamma_{dB}} = 5dB$.

At the first time slot, the SUs have not sensed any channels, and the myopic policy chooses the one with the highest stationary probability. All channels are characterized by (13) and have the same stationary probabilities. Since ties are resolved arbitrarily, the myopic policy has the same performance as the random strategy while the CSI-aided myopic policy achieves much better throughput early in the sensing process by preferring high-quality channels.

As time increases, the SUs learn from the instantaneous PU state and update the belief vectors as in (2). Thus the throughputs of the conventional and the CSI-aided myopic policies in Fig. 3 first improve, then decrease due to SU collisions, and finally approach their asymptotic values. However, the conventional myopic policy does not improve significantly on the random strategy due to SU congestion. In the randomized myopic policy, the chances for sensing SUs' favorite channels are reduced, and the individual throughput is sacrificed to compensate for possible SU collisions. Thus, performance of the randomized myopic policy is degraded in this scenario since SU competition is not extremely severe. The myopic/CA policy effectively reduces SU collisions and successfully restores the gain obtained from PU traffic cognition. However, it takes actions only after SUs collide, whereas the CSI-aided myopic policy takes advantage of the CSI variation to randomize the sensing decisions prior to sensing. Moreover, since the CSI-aided myopic policy selects stronger channels, it boosts the individual throughput. *The gain of the proposed policy over the myopic and myopic/CA strategies is about 1 and 0.8 bits per slot per user, respectively. Since all policies employ adaptive transmission, this gain is due to channel adaptation prior to sensing.*

While in Fig. 3 the fading channel is time-variant, we found that the throughput of the CSI-aided strategy is insensitive to the Doppler shift or slot duration although these parameters will affect the CSI estimation accuracy. Therefore, we assume the fading SNR $\gamma$ is fixed over 20 slots in the remaining numerical results and focus on the effect of channel mismatch in section IV.D below.

### B. Robustness to Spatial Correlation

In this section we explore adaptation to log-normal shadow fading where short-term fading is removed using diversity techniques. We investigate short-term (i.i.d) Rayleigh fading model and the long-term shadow fading models separately for the following reasons [23]. At lower speeds, log-normal shadowing remains almost constant for the duration of transmission, and Rayleigh fading changes sufficiently slowly to allow prediction and adaptation prior to sensing using collected pilots [13]. As the speed increases, prediction of the Rayleigh fading component might become infeasible, but it is still beneficial to adapt to shadow fading [23]. Thus, estimation and tracking of shadow fading CSI is simpler and more practical than for short-term fading CSI.

However, the shadow fading components of different SUs are likely to be correlated. To analyze the impact of spatial correlation on the performance of the proposed policy, we employ the correlated lognormal shadowing model [14] for a simple CR network topology with one common SU receiver and $M=20$ equally spaced SU transmitters placed uniformly on a linear track with minimum separation $d$ as shown in Fig. 4. The shadow fading coefficients are uncorrelated across different channels. For each channel, the correlation coefficient between any two links $m$ and $m'$ is given by [24]

$$\rho_{mm'} = \rho^{|m-m'|} \qquad (14)$$

with

$$\rho = e^{-ad}, \qquad (15)$$

where $a$ is a constant that depends on the environment. According to [24], $a \approx 0.12$ and $a \approx 0.002$ in urban and suburban cellular radio scenarios, respectively.

The impact of different values of $\rho$ is shown in Fig. 5. Each link is modeled using the lognormal distribution with average dB-scale SNR $\mu_{\gamma_{dB}} = 10 dB$ and the dB-spread $\sigma_{\gamma_{dB}} = 5 dB$. In this figure the throughput is averaged over $T = 20$ time slots. Perfect CSI and sensing results are assumed.

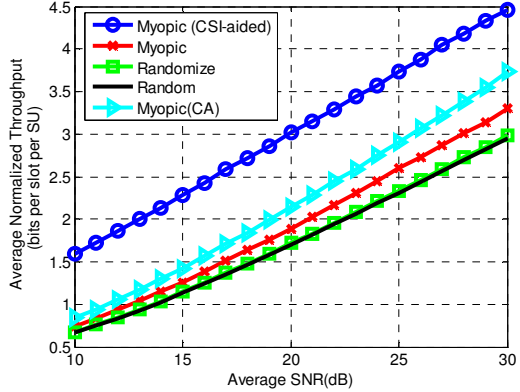

Fig. 6: Throughput vs. average SNR; continuous rate adaptation; fixed power; target BER=$10^{-3}$; 20 SU pairs; 40 channels; i.i.d Rayleigh fading.

Note that the throughput of the CSI-aided sensing strategy degrades as $\rho$ increases. For small $\rho$, the CSI-aided myopic policy randomizes decisions and benefits from multiuser diversity. However, when the SU links are highly correlated, i.e., for $\rho \geq 0.9$, the performance of the proposed CSI-aided sensing policy approaches that of the myopic policy. In this case, all SU links actually experience almost the same shadow fading patterns, so the multiuser diversity gained from adapting to the channel conditions is lost. From (15), $\rho \geq 0.9$ corresponds to the case when $d \leq 0.88 m$ in urban environments or $d \leq 52.68 m$ in suburban environments. However, the model in [14] is applicable to cellular scenarios and does not accurately characterize channel propagation in vehicle-to-vehicle ad hoc networks. Spatial correlation among ad hoc network links was investigated in [15] and shown not to exceed $\rho = 0.3$ in most scenarios. *The proposed CSI-aided policy maintains most of its potential gain for such correlation values.*

### C. Adaptive Modulation

Next, we investigate the throughput averaged over 20 slots for the sensing strategies that employ *adaptive modulation* reward (5) in Fig. 6 as a function of the average SNR. All other assumptions are the same as in section IV.A above.

We observe that the proposed CSI-aided policy outperforms the myopic policy by at least 8 dB. The gain over the myopic/CA policy [11] decreases with SNR, but is *at least 5dB* for practical SNR $\leq 30dB$. Similar gains were observed when the capacity reward is employed and for the shadow fading channel model.

### D. Robustness to Imperfect CSI and Sensing Errors

In Fig. 7 we study robustness of the CSI-aided myopic policy to *CSI mismatch* over i.i.d. Rayleigh fading channel for a smaller CR network with $M = 3$ SU pairs and $N = 10$ channels. We use the capacity reward given mismatched observation (7), where $R(\gamma)$ is given by (4), and $f(\gamma | \hat{\gamma})$ is computed as in (8). Since CSI estimation during channel exploration is likely to be more accurate than prior to sensing, we assume perfect CSI during adaptive transmission for all strategies and focus on the impact of the CSI error on the performance of channel-adaptive sensing. We observe that the throughput of the CSI-aided sensing policy provides multiuser diversity with mismatched CSI and approximates the ideal CSI case when NMSE $\sigma^2 < 0.1$. Note that $\sigma^2 = 0.1$ corresponds to severely degraded CSI prediction accuracy in conventional adaptive transmission applications [13]. The throughput of the CSI-aided strategy converges to that of the conventional myopic policy as the NMSE $\sigma^2$ approaches 1, i.e., when the CSI information is not available. *Thus the proposed approach is robust to CSI errors and degrades gracefully when CSI becomes unreliable.*

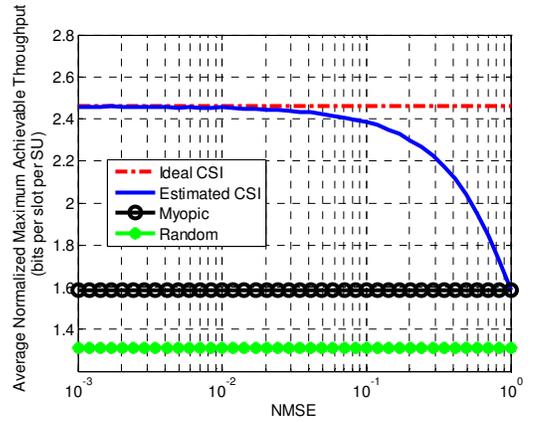

Fig. 7: Throughput vs. NMSE prior to sensing under CSI error; capacity reward; 3 SU pairs; 10 channels; i.i.d Rayleigh fading; average SNR=10dB.

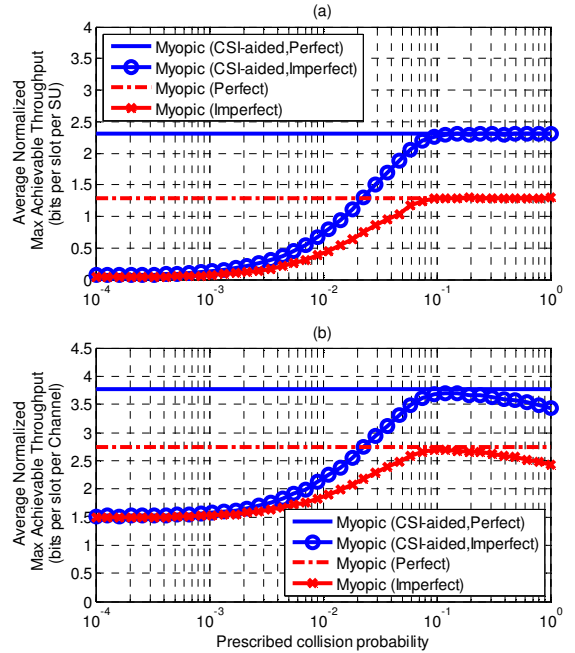

Fig. 8: Throughput vs. $p_m$ under sensing error; capacity reward; i.i.d Rayleigh fading; 20 SU pairs; 40 channels; average SNR=10dB; $\nu = 5$.

Finally, we investigate the effect of *sensing errors* assuming energy detection is the underlying spectrum sensing method at the physical layer. We assume all links in the network are i.i.d. Rayleigh fading with average SNR=10dB. The number of samples is $\nu = 5$, and the belief update is specified in (9-12).

In Fig. 8, the throughput vs. allowed collision probability is compared for the proposed CSI-aided myopic strategy and the conventional myopic policy. The average normalized maximum achievable rate *of the SU network* is illustrated in (a), and the overall throughput (normalized by the number of channels) for *both PU and CR networks* is shown in (b). As we increase the allowed collision probability with the PU network, the average throughput of the SUs increases and eventually saturates to the ideal spectrum sensor case for each sensing strategy. However, for large $p_m$, the PU transmission is frequently interrupted by the SUs, causing the overall spectral efficiency to decline. For both policies, the SU network can approximate the ideal throughput in the presence of sensing error unless the collision constraint is extremely tight. In this scenario, the maximum spectrum efficiency is achieved by setting $p_m \approx 0.1$. We observe that *the proposed CSI-aided strategy maintains about 1 bit gain over the conventional myopic strategy for $p_m \geq 0.1$.*

## V. CONCLUSION AND FUTURE WORK

A novel channel-adaptive sensing strategy was proposed for ad hoc CR networks where neighboring SUs experience similar spectrum opportunities. This strategy randomizes sensing decisions and boosts the throughput by adapting the reward to the channel CSI. It was demonstrated that the proposed sensing strategy achieves a gain of about 1 bit per slot per user relative to previously investigated sensing approaches in a medium size moderately congested network. Since all compared strategies employ adaptive transmission, this gain is due to adaptation to channel CSI prior to sensing. Moreover, the proposed strategy is robust to CSI mismatch, spatial correlation among SU links, and sensing errors.

We found that the proposed policy does not completely eliminate secondary collisions. Therefore, we plan to combine CSI-aware sensing with interference mitigation algorithms to resolve SU congestion for diverse network scenarios. We will also address the design and trade-offs of pilot placement and CSI prediction to enable channel-aware sensing in realistic ad hoc networks. Modeling of temporal, spatial, and frequency correlation under shadow fading among ad hoc network links will be employed to further verify robustness of CSI-aided sensing strategies. Finally, in practice PU traffic is time-variant, so we will explore sensing strategies that track and adapt to PU traffic statistics as well as to channel CSI.